\documentclass[12pt, a4paper]{article}
\usepackage[dvipdfmx]{graphicx}
\usepackage{amssymb}
\usepackage{amsmath}
\usepackage{amsthm}
\usepackage{bm}
\usepackage{cite}
\usepackage{color}
\usepackage{cancel}

\usepackage[dvipdfm]{hyperref}

\begin{document}

\begin{titlepage}

\begin{flushright}

IPMU 13-0165

\end{flushright}

\vskip 1.35cm
\begin{center}

{\large 
{\bf 
QCD Corrections to Quark (Chromo)Electric Dipole Moments \\
in High-scale Supersymmetry 
}
}

\vskip 1.2cm

Kaori Fuyuto$^a$,
Junji Hisano$^{a,b}$,
Natsumi Nagata$^{a,c}$,
and
Koji Tsumura$^{a}$

\vskip 0.4cm

{\it $^a$Department of Physics,
Nagoya University, Nagoya 464-8602, Japan}\\
{\it $^b$Kavli Institute for the Physics and Mathematics of the Universe
 (WPI), Todai Institutes for Advanced Study, the University of Tokyo,
 Kashiwa 277-8568, Japan}\\
{\it $^c$Department of Physics, 
the University of Tokyo, Tokyo 113-0033, Japan}
\date{\today}

\vskip 1.5cm

\begin{abstract} 

 Recent results from the LHC experiments, both for the Higgs mass
 measurement and the direct search for supersymmetric (SUSY) particles,
 might indicate that the SUSY breaking scale is much higher than the
 electroweak scale. Although it is difficult to investigate such a
 scenario at collider experiments, the measurement of the hadronic electric
 dipole moments is one of promising ways to detect the effects of the SUSY
 particles. These effects are expressed in terms of the CP-violating
 effective operators defined at the SUSY breaking scale, which involve
 quarks, gluons, photons, and gluinos. In this paper, we discuss the QCD
 corrections to the effective operators in the high-scale SUSY
 scenario. To appropriately evaluate the radiative corrections in the
 presence of large mass hierarchy among the SUSY particles, we exploit
 an effective theoretical approach based on the renormalization-group
 equations.  As a result, it is found that the low-energy quark electric
 and chromoelectric dipole moments may differ from those evaluated in
 previous works by ${\cal O}(100)$ \% and ${\cal O}(10)$ \%,
 respectively.

\end{abstract}

\end{center}
\end{titlepage}

\section{Introduction}
\label{intro}

The supersymmetric (SUSY) extension of the Standard Model (SM) is a
leading candidate for physics beyond the SM. So far, however, the
weak-scale SUSY models have been severely restricted since
no evidence for new physics has been found yet; for
instance, the latest results from the LHC experiments have imposed
stringent limits on the masses of the SUSY particles, especially those
of colored particles \cite{ATLAS-CONF-2013-047, Chatrchyan:2013lya}. In
addition, the Higgs boson with a mass of $\sim 126$~GeV \cite{PDG2013},
which was recently discovered at the LHC~\cite{:2012gk, :2012gu}, might
also indicate that the SUSY particles are well above ${\cal O}(1)$~TeV,
since in the minimal supersymmetric Standard Model (MSSM) sufficient
radiative corrections are required in order to realize the mass of the
Higgs boson \cite{Okada:1990vk, Okada:1990gg, Haber:1990aw, Ellis:1990nz,
Ellis:1991zd}. Unless the Higgs sector is modified nor the left- and 
right-handed stops adequately mix with each other, such a large quantum
effect is only provided with heavy stops having masses of much higher
than the electroweak scale.

The current situation motivates us to study models with a high
SUSY breaking scale. Such models assume that SUSY is broken at a scale of
${\cal O}(10^{2-3})$~TeV to yield the 126~GeV Higgs boson
\cite{Giudice:2011cg, Ibe:2011aa, Ibe:2012hu, Ibanez:2013gf}. In this
case, scalar particles except the lightest Higgs boson acquire masses of
the order of the SUSY breaking scale. Fermionic superpartners, on the
other hand, may be much lighter than the other sparticles since their
masses are protected by chiral symmetries. Indeed, such a mass spectrum
is realized with a simple SUSY breaking mechanism in which SUSY is broken
by a non-singlet field and the breaking effects are transmitted to the
visible sector via a generic K\"{a}hler potential. In this framework, the
gaugino masses are induced by the anomaly mediation
\cite{Randall:1998uk, Giudice:1998xp} and suppressed by one-loop factors
compared with the scalar mass. With the SUSY
breaking scale being ${\cal O}(10^{2-3})$~TeV, gauginos may lie around
the TeV scale. The neutral wino turns out to be the lightest SUSY
particle in this model, and may make up a main component of the dark matter
in the Universe \cite{Gherghetta:1999sw, Moroi:1999zb,
Hisano:2006nn}. Further, it is found that the gauge coupling unification
is not only preserved but improved in the scenario
\cite{Hisano:2013cqa}. Thus, the high-scale SUSY models have interesting
features from a phenomenological point of view \cite{Wells:2003tf,
ArkaniHamed:2004fb, Giudice:2004tc, ArkaniHamed:2004yi, Wells:2004di,
Hall:2011jd}, and recently attract a lot of attention especially after
the early LHC runnings \cite{Jeong:2011sg, Saito:2012bb,
Bhattacherjee:2012ed, Arvanitaki:2012ps, Hall:2012zp, Hisano:2012wm,
ArkaniHamed:2012gw, Hisano:2013exa}.

Although it is difficult to investigate the high-scale SUSY scenario at high-energy
collider experiments, the low-energy precision experiments might catch up
the SUSY signature. Without any flavor symmetries, soft SUSY breaking
parameters in general give rise to extra sources of flavor and/or CP
violation \cite{Gabbiani:1996hi}. These effects, such as the
flavor-changing neutral currents and the hadronic and leptonic electric dipole moments (EDMs), are
suppressed by sfermion masses, and thus the high-scale SUSY models do not
conflict with the current experimental results for these quantities. 
Among such experiments, the measurement of the EDMs offers 
a promising way to look for the signature of the SUSY
particles \cite{Moroi:2013sfa, McKeen:2013dma}. Since in the SM
the EDMs induced by the CP phase in the Cabibbo-Kobayashi-Maskawa (CKM) matrix 
are considerably below the sensitivities of the present and near future experiments
\cite{Mannel:2012qk, Khriplovich:1981ca}, the EDM measurement
is free from the SM background, thus provides a clean environment to
detect a sign of high-energy physics beyond the SM.

The effective interactions which give rise to the EDMs are
expressed in terms of the flavor-conserving CP-violating effective
operators.  In SUSY
models, such operators are induced by diagrams in which SUSY particles
run in the loop. In the case of the high-scale SUSY scenario, however, one
needs to pay particular attention to the calculation of the diagrams;
as mentioned above, there exists a large difference between the mass
scales of scalar and fermionic SUSY particles and this hierarchy causes 
large logarithmic factors which may spoil the perturbation theory. To
evade the difficulties, we need to evaluate the effective operators by
means of the renormalization-group equations (RGEs).
The renormalization corrections are particularly important for the operators
including colored particles because of the large value of the strong
coupling constant. 

In this paper, we study the QCD effects on
the flavor-preserving CP-odd quark and gluon operators generated by the
squark-gluino interactions. Among the operators, the EDMs and the
chromoelectric dipole moments (CEDMs) of quarks have the lowest
mass-dimensions, and thus sensitive to the SUSY contribution. We focus
on these two operators and study the contribution of the quark-gluino
four-Fermi operators to the
quantities. The calculation is divided into two steps; first, by
integrating out squarks, we construct an effective theory with quarks,
gluons, photons, and gluinos. Then, the effective operators are evolved
down to the gluino threshold according to the RGEs. 
During the RGE flow, the CEDMs are radiatively generated from the
dimension-six quark-gluino operators.
The resultant EDMs
and CEDMs evaluated in this way are compared with the 
results based on the computation of the one-loop diagrams. Possible
ways of improvement of the calculation are also discussed.

This paper is organized as follows. In the next section, we write down
the CP-violating effective operators involving quarks, gluons, photons,
and gluinos which we consider in the following discussion, and present
the anomalous dimension matrix for the operators. The Wilson
coefficients of the effective operators are evaluated in the MSSM with
and without the assumption of the minimal flavor violation in the
sections \ref{MFV} and \ref{generic}, respectively. Evolving them down
according to the RGEs, we obtain the EDMs and CEDMs of light quarks at
the hadron scale, and compare them with the explicit one-loop
calculation. Section~\ref{conclusion} is devoted to conclusion and
discussion.

\section{Effective Lagrangian}

To begin with, we write down the CP-violating effective operators at the
hadron scale ($\sim 1$~GeV)
which consist of the flavor-diagonal operators of light quarks, photons,
and gluons up to dimension-five:
\begin{align}
 {\cal L}_{\tiny \cancel{\rm {CP}}}&=-\sum_{q=u,d,s}m_q
 \bar{q} i\theta_q \gamma_5q + \theta_G
 \frac{\alpha_s}{8\pi}G^A_{\mu\nu}\widetilde{G}^{A\mu\nu} \nonumber \\
&-\frac{i}{2}\sum_{q=u,d,s}d_q\bar{q}\sigma^{\mu\nu}\gamma_5qF_{\mu\nu} 
-\frac{i}{2}\sum_{q=u,d,s}\tilde{d}_q\bar{q}g_s\sigma^{\mu\nu}\gamma_5T^Aq
G^A_{\mu\nu}  
~.
\label{Lagrangian}
\end{align}
Here, $m_q$ are the quark masses, $g_s$ is the strong coupling constant
($\alpha_s=g_s^2/4\pi)$, and $T^A$ are the generators of the SU(3)$_{\rm
C}$. $F_{\mu\nu}$ and $G^A_{\mu\nu}$ are the field strength tensors of
photon and gluon, and their dual fields are defined by, {\it e.g.},
$\widetilde{G}^A_{\mu\nu}\equiv
\frac{1}{2}\epsilon_{\mu\nu\rho\sigma}G^{A\rho\sigma}$ with
$\epsilon^{0123}=+1$. The second term of the above expression is
the effective QCD $\theta$ term, which is connected with the first term
through the chiral rotation. These two terms are suppressed in the
presence of the Peccei-Quinn symmetry \cite{Peccei:1977hh}.
The third and fourth terms represent the EDMs and the CEDMs for light
quarks, respectively. They are dimension-five operators, and thus quite
sensitive to the high-scale 
physics beyond the SM.

Apart from the SM contribution, these operators are induced by
diagrams where SUSY particles run in the loop. In this paper, we focus
on the SUSY contribution and discuss the QCD effects on it at the
leading order in $\alpha_s$. In
particular, we consider the case where a large mass
difference between the scalar particles and gauginos exists. As mentioned
in the Introduction, such a hierarchical mass spectrum often shows up
in the high-scale SUSY models. To appropriately include the QCD corrections
in the presence of the large mass hierarchy, we evaluate them based on
the method of the effective field theory as well as the RGEs. First, by
integrating out squark fields, we obtain the effective Lagrangian below
the SUSY breaking scale, which involves only gluinos and the SM
fields. The short-distance effects, which reflect the CP-violation due
to the SUSY particles, are included into the Wilson coefficients of the
effective operators matched at the SUSY breaking scale. Next, the
effective operators are evolved down to the gluino threshold according to
the RGEs. Then, at the threshold, the gluino fields are integrated out
to give the effective theory which contains only the SM fields. After
this step, the ordinary procedure is applied to estimate the effects of
the CP-violating operators on the low-energy physics such as the neutron
EDM. The purpose of this paper is to formulate the first two steps in
terms of the operator product expansions and the RGEs. 

The effective Lagrangian below the SUSY breaking scale is given as
follows:  
\begin{equation}
 {\cal L}_{\rm eff}=\sum_{q=u,d,s}C_1^q(\mu){\cal
  O}^q_1(\mu)+\sum_{q=u,d,s}C_2^q (\mu){\cal O}^q_2(\mu)
+\sum_{q=u,d,s}\sum_{i=1}^{5}
\widetilde{C}^q_i(\mu)\widetilde{\cal G}^q_i(\mu)~,
\label{eff_Lag}
\end{equation}
where
\begin{align}
 {\cal O}^q_1&\equiv -\frac{i}{2}eQ_qm_q\overline{q}\sigma^{\mu\nu}
\gamma_5 qF_{\mu\nu}~,\nonumber\\
 {\cal O}^q_2&\equiv -\frac{i}{2}g_sm_q\overline{q}\sigma^{\mu\nu}
\gamma_5 T^AqG^A_{\mu\nu}~,\nonumber\\
\widetilde{\cal G}^q_1&\equiv
 \frac{1}{2}\bar{q}q\overline{\tilde{g}} {}^Ai\gamma_5 \tilde{g}^A~,
\nonumber\\
\widetilde{\cal G}^q_2&\equiv
 \frac{1}{2}\bar{q}i\gamma_5q\overline{\tilde{g}}{}^A \tilde{g}^{A}~,
\nonumber\\
\widetilde{\cal G}^q_3&\equiv
 \frac{1}{2}d_{ABC}\bar{q}T^Aq\overline{\tilde{g}}{}^Bi\gamma_5
 \tilde{g}^C~,\nonumber\\
\widetilde{\cal G}^q_4&\equiv
 \frac{1}{2}d_{ABC}\bar{q}i\gamma_5T^Aq\overline{\tilde{g}}{}^B\tilde{g}^C~,
\nonumber\\
\widetilde{\cal G}^q_5&\equiv
 \frac{i}{2}f_{ABC}\bar{q}\sigma^{\mu\nu}i\gamma_5T^Aq
\overline{\tilde{g}}{}^B
\sigma_{\mu\nu} \tilde{g}^C~.
\end{align}
Here, $Q_q$ are the electric charges for light quarks with
$(Q_u,Q_d,Q_s)=(2/3,-1/3,-1/3)$. The covariant derivative for quarks is
defined as $D_\mu \equiv \partial_\mu -ieQ_qA_\mu-ig_s G^A_\mu T^A$
($e<0$) with $A_\mu$ and $G^A_\mu$ the U(1)$_{\rm EM}$ and SU(3)$_{\rm
C}$ gauge fields, respectively. $\tilde{g}^A$ denotes gluinos, which are
Majorana fermions and form an adjoint representation under the
SU(3)$_{\rm C}$ transformations. The totally symmetric factor $d_{ABC}$
is defined by $d_{ABC}\equiv 2{\rm Tr}(\{T_A, T_B\}T_C)$, while
$f_{ABC}$ is the structure constant of the SU(3) group with $[T_A,
T_B]=if_{ABC}T_C$. The Wilson coefficients of the operators, $C^q_1$,
$C^q_2$, and $\widetilde{C}^q_i$, are obtained by integrating out squark
fields at the SUSY breaking scale. 

In Eq.~\eqref{eff_Lag} we only keep the operators which give significant
corrections to the quark EDMs and CEDMs. 
Let us comment on the operators which we ignore in the following
analysis. First, we do not consider the dimension-four operators in
Eq.~\eqref{Lagrangian} since they do not contribute to the RGEs for the
dimension-five operators. Especially, when the Peccei-Quinn symmetry is
imposed, these dimension-four operators are suppressed and the EDMs and
CEDMs give dominant contributions to the hadronic and atomic EDMs. Also,
we ignore the dimension-five gluino CEDM,
$f_{ABC}g_s\overline{\tilde{g}}{}^A\sigma^{\mu\nu}\gamma_5
\tilde{g}^BG^C_{\mu\nu}$, since it does not affect the running of the
operators in Eq.~\eqref{eff_Lag} at the leading order in $\alpha_s$. As
for the dimension-six operators, the Weinberg operator
$f_{ABC}G^A_{\mu\nu} \widetilde{G}^{B\nu\lambda}G^C_\lambda{}^{\mu}$
\cite{Weinberg:1989dx}, four-quark operators, and four-gluino operators
might also yield sizable effects on the radiative corrections to the
operators above. These dimension-six operators are, however, generated
at ${\cal O}(\alpha_s^2)$ in the case of the MSSM, and thus safely
neglected in the leading order calculation.
\footnote{
Note that once these dimension-six operators are induced, they actually
give rise to significant contributions to the EDMs and the CEDMs. For
instance, at one-loop level, the Weinberg operator mixes with the quark
CEDMs \cite{Braaten:1990gq}, while four-quark operators including heavy
quarks radiatively induce both the EDMs and the CEDMs
\cite{Hisano:2012cc}. 
}

Next, we evaluate the anomalous dimensions of the operators in
Eq.~\eqref{eff_Lag} at the leading order. The RGE for the Wilson
coefficients in Eq.~\eqref{eff_Lag} is written as
\begin{equation}
 \mu \frac{\partial}{\partial \mu}\Vec{C}(\mu)
=\Vec{C}(\mu)\Gamma~,
\label{rge}
\end{equation}
where $\Vec{C}$ is a column vector defined by
\begin{equation}
 \Vec{C}\equiv
  (C^q_1,C^q_2,\widetilde{C}^q_1,\widetilde{C}^q_2,
\widetilde{C}^q_3,\widetilde{C}^q_4,\widetilde{C}^q_5)  ~.
\end{equation}
Then, we obtain the following anomalous dimension matrix at one-loop
level:
\begin{equation}
 \Gamma =
\begin{pmatrix}
 \frac{\alpha_s}{4\pi}\gamma_q&0\\[10pt]
 \frac{1}{(4\pi)^2}\gamma_{q\tilde{g}}&
 \frac{\alpha_s}{4\pi}\gamma_{\tilde{g}} 
\end{pmatrix}
~,
\end{equation}
with
\begin{equation}
 \gamma_q=
\begin{pmatrix}
 8C_F&0\\
8C_F&16C_F-4N
\end{pmatrix}
~,
\end{equation}
\begin{equation}
 \gamma_{\tilde{g}}=
\begin{pmatrix}
 -6C_F-6N&0&0&0&2\\
0&-6C_F-6N&0&0&2\\
0&0&-6C_F&0&(N^2-4)/2N\\
0&0&0&-6C_F&(N^2-4)/2N\\
24&24&12N&12N&2C_F-4N
\end{pmatrix}
~,
\label{gammag}
\end{equation}
and
\begin{equation}
 \gamma_{q\tilde{g}}
=
\begin{pmatrix}
 0&0\\
 0&0\\
 0&0\\
 0&0\\
0&8N\frac{M_{\tilde{g}}}{m_q}
\end{pmatrix}
~.
\label{mixing}
\end{equation}
Here, $N(=3)$ is the number of colors, $C_F=(N^2-1)/2N$ is the
quadratic Casimir invariant for the fundamental representation, and
$M_{\tilde{g}}$ is the mass of gluino. The anomalous dimension matrix
for the dimension-five operators $\gamma_q$ is readily
obtained from that for the dipole-type operators relevant to the $b\to
s\gamma$ process \cite{Shifman:1976de, Ciuchini:1993fk}. Note that the
coefficient of $\gamma_{q\tilde{g}}$ is not suppressed by the strong
coupling constant $\alpha_s$.
In this case, the scale-dependence arises from a mismatch in the
dimension between the dimension-five and -six operators
\cite{Grinstein:1990tj}. 
 A similar feature is found in the case of
four-quark operators mixing into the quark EDMs and CEDMs, as discussed
in Ref.~\cite{Hisano:2012cc}.

\section{MSSM with minimal flavor violation}
\label{MFV}

Now all we have to do is to compute the Wilson coefficients in
Eq.~\eqref{eff_Lag} by integrating out squarks in a certain model. Then,
by evolving them down according to the RGE \eqref{rge}, we obtain the
quark EDMs and CEDMs in the low-energy region. In the following
discussion, we take up the MSSM as an example. Also, in this section, we
focus on the case with the so-called minimal flavor violation
\cite{Hall:1990ac, D'Ambrosio:2002ex}, which assumes that the CKM matrix
is the only source for all of the flavor-violating terms in the MSSM.

In the present case, the tree-level squark exchanging diagrams give rise
to the CP-odd quark-gluino four-Fermi operators, which are to induce the
quark CEDMs radiatively. The squark mass matrix has the following form:
\begin{equation}
 {\cal L}_{\rm mass}=-
\begin{pmatrix}
 \tilde{q}{}^*_L &\tilde{q}{}^*_R
\end{pmatrix}
\begin{pmatrix}
 m^2_{\tilde{q}^{}_L}&m_qX_q\\[6pt]
m_qX^*_q& m^2_{\tilde{q}^{}_R}
\end{pmatrix}
\begin{pmatrix}
 \tilde{q}^{}_L\\[6pt] \tilde{q}^{}_R
\end{pmatrix}
~,
\end{equation}
where $\tilde{q}^{}_L$ and $\tilde{q}^{}_R$ represent the left- and
right-handed squarks, respectively, and $X_u\equiv A^*_u-\mu \cot\beta$
($X_d\equiv A^*_d-\mu \tan\beta$) for up-type (down-type) quarks. Here
we assume that the trilinear soft scalar couplings (the so-called
$A$-terms) $A_q$ are proportional to the corresponding Yukawa couplings. 
In $X_q$, $\mu$ is the higgsino-mass parameter and $\tan\beta$ is the
ratio of the vacuum expectation values of the MSSM Higgs fields. 
Throughout this article we take a convention where the gaugino masses
are set to be real parameters, without loss of generality. On the
assumption of the minimal flavor violation, the flavor-mixings in the
squark mass matrix are considerably suppressed, so we neglect them in the
present discussion. We also take $m^2_{\tilde{q}^{}_L}=
m^2_{\tilde{q}^{}_R}= M_S^2$ and $m_qX_q\ll M_S^2$, for simplicity. 
Then, by evaluating the squark exchanging diagrams, we readily obtain
the Wilson coefficients at the scalar mass scale $M_S$:
\begin{equation}
  C^q_1(M_S)=C^q_2(M_S)=0~,
\label{initial0}
\end{equation}
and
\begin{align}
 \widetilde{C}^q_1(M_S)= \widetilde{C}^q_2(M_S)&=
-\frac{1}{2N}\frac{g_s^2m_q}{M_S^4}{\rm
 Im}(X_q) ~,\nonumber\\
 \widetilde{C}^q_3(M_S)= \widetilde{C}^q_4(M_S)&=
-\frac{1}{2}\frac{g_s^2m_q}{M_S^4}{\rm
 Im}(X_q) ~,\nonumber\\
 \widetilde{C}^q_5(M_S)&=+\frac{1}{4}\frac{g_s^2m_q}{M_S^4}{\rm
 Im}(X_q) ~.
 \label{initial}
\end{align}
Notice that the quark EDMs and CEDMs vanish at tree-level. They
  are induced radiatively through the mixing terms in RGEs
  and also from the short-distance contribution,
  as will be shown below.

By using Eqs.~\eqref{initial0} and \eqref{initial} as initial
conditions, we solve the RGE \eqref{rge} to evaluate the Wilson coefficients at the gluino
threshold. Especially, in the leading-logarithmic approximation, the
quark CEDMs are generated as
\begin{equation}
C^q_2(M_{\tilde{g}})
\simeq-\frac{1}{(4\pi)^2}8N\frac{M_{\tilde{g}}}{m_q} 
\ln\biggl(\frac{M_S}{M_{\tilde{g}}}\biggr)
~\widetilde{C}^q_5(M_S)
~,
\label{cedmll}
\end{equation}
while the EDMs vanish at the leading order. This result is to be
compared with the explicit calculation of the one-loop gluino-squark
diagrams. In the limit of $M_{\tilde{g}}\ll M_S$, we have
\cite{Degrassi:2005zd} 
\begin{align}
C^q_1\vert_{\rm 1loop}&= +\frac{1}{(4\pi)^2}
 \frac{16}{3}\frac{M_{\tilde{g}}}{m_q}~ \widetilde{C}^q_5(M_S)~,
\label{edm1loop} \\[8pt]
C^q_2\vert_{\rm 1loop}&= -\frac{1}{(4\pi)^2}
\biggl[
8N\ln\biggl(
\frac{M_S}{M_{\tilde{g}}}
\biggr)-\frac{88}{3}
\biggr]\frac{M_{\tilde{g}}}{m_q}
~ \widetilde{C}^q_5(M_S)~.
\label{cedm1loop}
\end{align}
The first term in Eq.~\eqref{cedm1loop} is consistent with
Eq.~\eqref{cedmll}. The non-logarithmic terms in Eqs.~\eqref{edm1loop} and
\eqref{cedm1loop} result from the short-distance contribution; it is
induced by the processes in which the loop integrals are dominated by
momenta around $M_S$. In that sense, the first term in
Eq.~\eqref{cedm1loop} is to be regarded as the long-distance
contribution, with the factorization scale around the squark mass
scale.

\begin{figure}[t]
\begin{center}
\includegraphics[width=70mm]{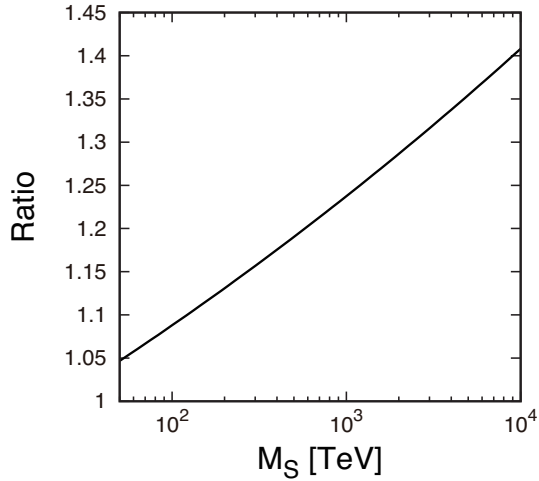}
\caption{Ratio $C^q_2(M_{\tilde g})/C^q_2|^{(L)}_{\rm 1loop}$
 against the squark mass $M_S$. Gluino mass is fixed to $M_{\tilde g}=3
 \ {\rm TeV}$.} 
\label{LOratioCEDM}
\end{center}
\end{figure}

To see the significance of the running effects, we evaluate $C^q_2$ at
the gluino threshold numerically and compare it with the long-distance
part of Eq.~\eqref{cedm1loop}, {\it i.e.},
\begin{equation}
 C^q_2\vert_{\rm 1loop}^{(L)}
= -\frac{1}{(4\pi)^2}8N\frac{M_{\tilde{g}}}{m_q}
\ln\biggl(
\frac{M_S}{M_{\tilde{g}}}
\biggr)
~ \widetilde{C}^q_5(M_S)~.
\end{equation}
The difference is caused by the running of the parameters and
the mixing among the effective operators. In Fig.~\ref{LOratioCEDM}, we plot
the ratio $C^q_2(M_{\tilde{g}})/C^q_2\vert_{\rm 1loop}^{(L)}$
against the squark mass $M_S$. Here, the gluino mass is fixed to
$M_{\tilde g}=3 \ {\rm TeV}$. In $C^q_2\vert_{\rm 1loop}^{(L)}$ and
$\widetilde{C}^q_5(M_S)$, we use $M_{\tilde{g}}$ and $m_q$ evaluated at
the squark mass scale. Moreover, in order to obtain
$C^q_2(M_{\tilde{g}})$, the RGEs are solved using the beta function of
the strong coupling constant which contains the contribution of both
gluino and SM particles. Figure~\ref{LOratioCEDM} shows that as
the squark mass scale becomes large, the running-effects yield the
${\cal O}(10) \% $ difference between $C^q_2(M_{\tilde{g}})$ and
$C^q_2\vert_{\rm 1loop}^{(L)}$.

\begin{figure}
\begin{minipage}{0.5 \hsize}
\begin{center}
\includegraphics[width=70mm]{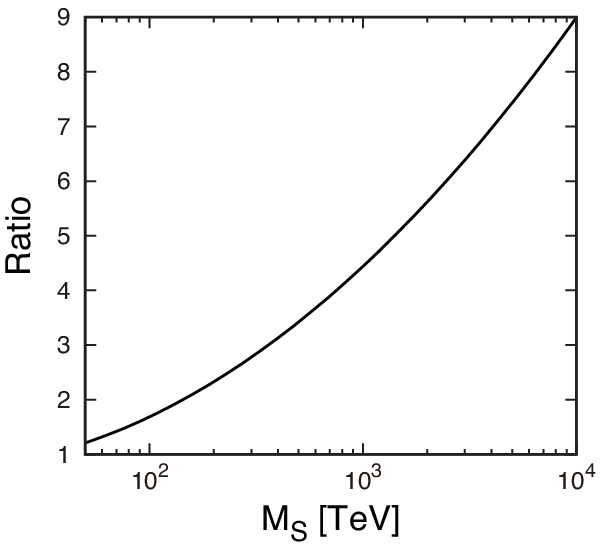}
\end{center}
\end{minipage}
\begin{minipage}{0.5 \hsize}
\begin{center}
\includegraphics[width=70mm]{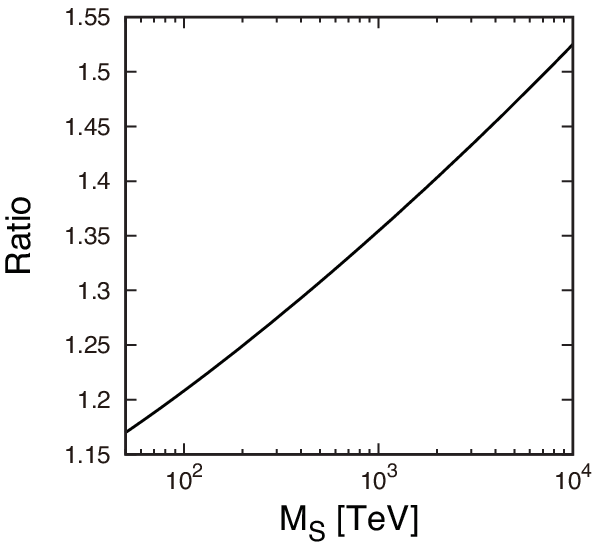}
\end{center}
\end{minipage}
\caption{Ratios $C^q_1(M_{\tilde{g}})/C^q_1\vert_{\rm
 1loop}$ and $C^q_2(M_{\tilde{g}})/C^q_2\vert_{\rm 1loop}$ as functions
 of $M_S$ in left and right graphs, respectively. In both graphs, gluino
 mass is fixed to $M_{\tilde{g}} = 3~{\rm TeV}$.}  
\label{RatioWshort}
\end{figure}

Next, we take the threshold short-distance contributions into account, and
evaluate both $C^q_1(M_{\tilde{g}})$ and $C^q_2(M_{\tilde{g}})$ in terms
of the RGEs. Then, they are compared with the explicit one-loop results in
Eqs.~\eqref{edm1loop} and \eqref{cedm1loop}. The initial conditions for
$C^q_1$ and $C^q_2$ are given by the short-distance contribution in
Eqs.~\eqref{edm1loop} and \eqref{cedm1loop}, that is,
\begin{align}
 C^q_1(M_S)&= +\frac{1}{(4\pi)^2}
 \frac{16}{3}\frac{M_{\tilde{g}}}{m_q}
~ \widetilde{C}^q_5(M_S)~, \nonumber\\[6pt]
C^q_2(M_S)&= +\frac{1}{(4\pi)^2}
\frac{88}{3}\frac{M_{\tilde{g}}}{m_q}
~ \widetilde{C}^q_5(M_S)~,
\end{align}
while those for $\widetilde{C}^q_i$ ($i=1,\dots,5$) are given by
Eq.~\eqref{initial}. In Fig.~\ref{RatioWshort}, the results are plotted
as functions of $M_S$. Here again, the gluino mass is fixed to
$M_{\tilde g}=3 \ {\rm TeV}$. The left (right) panel in
Fig.~\ref{RatioWshort} represents the ratio
$C^q_1(M_{\tilde{g}})/C^q_1\vert_{\rm 1loop}$
($C^q_2(M_{\tilde{g}})/C^q_2\vert_{\rm 1loop}$). 
As for $C^q_2$, it is again found that the variation of the
squark mass scale may change the ratio by ${\cal O}(10) \%$. In
the case of $C^q_1$, on the other hand,  the RGE result is several times
larger than the explicit one-loop result, which is quite drastic
compared to the case of $C^q_2$. It is found that this enhancement
is caused by the mixing of the CEDM operators, whose contribution
becomes dominant as the squark mass scale taken to be higher.

\section{MSSM with a generic flavor structure}
\label{generic}

In the high-scale SUSY scenario, flavor-violation in the soft mass terms
of squarks is allowed to be sizable, which motivates us to consider the
case where squark mass matrices have a generic flavor structure. In such
a case the dominant contributions to the EDMs and CEDMs of light quarks
come from the flavor-violating processes \cite{Hisano:2004tf,
Hisano:2008hn}. These contributions are also evaluated with the
prescription described in the 
previous section. The Wilson coefficients of the effective operators at
the SUSY breaking scale in the present case are given as
\begin{align}
 \widetilde{C}^q_1(M_S)= \widetilde{C}^q_2(M_S)&=
-\frac{1}{2N}\frac{g_s^2m_{q_3^{}}}{M_S^4}{\rm
 Im}\bigl[
(\delta_{LL})_{qq_3^{}}X_{q_3^{}}(\delta_{RR})_{q_3^{}q}
\bigr] ~,\nonumber\\
 \widetilde{C}^q_3(M_S)= \widetilde{C}^q_4(M_S)&=
-\frac{1}{2}\frac{g_s^2m_{q_3^{}}}{M_S^4}{\rm
 Im}\bigl[
(\delta_{LL})_{qq_3^{}}X_{q_3^{}}(\delta_{RR})_{q_3^{}q}
\bigr] ~,\nonumber\\
 \widetilde{C}^q_5(M_S)&=+\frac{1}{4}\frac{g_s^2m_{q_3^{}}}{M_S^4}{\rm
 Im}\bigl[
(\delta_{LL})_{qq_3^{}}X_{q_3^{}}(\delta_{RR})_{q_3^{}q}
\bigr] ~,
 \label{initial_FV}
\end{align}
where $q_3^{}$ denotes $t$-quark ($b$-quark) for the up-type (down-type)
quarks, and the mass insertion parameters \cite{Gabbiani:1996hi,
Hall:1985dx} $(\delta_{LL})_{ij}$ and $(\delta_{RR})_{ij}$ are defined
by
\begin{equation}
 (\delta_{LL})_{ij}\equiv \frac{( m^2_{\tilde{q}_L})_{ij}}{M_S^2}~,
~~~~~~
 (\delta_{RR})_{ij}\equiv \frac{( m^2_{\tilde{q}_R})_{ij}}{M_S^2}~.
\end{equation}
It is possible for them to be ${\cal O}(1)$ in the high-scale SUSY
scenario \cite{McKeen:2013dma}. Thus, the above coefficients are
enhanced by the Yukawa coupling constants of the third generation quarks without
suffering from the suppression. In addition, we take into account the
short-distance threshold corrections at one-loop level for $C^q_1$ and
$C^q_2$: 
\begin{align}
 C^q_1(M_S)&= +\frac{1}{(4\pi)^2}
 \frac{16}{3}\frac{M_{\tilde{g}}}{m_q}
~ \widetilde{C}^q_5(M_S)~, \nonumber\\[6pt]
C^q_2(M_S)&= +\frac{1}{(4\pi)^2}
\frac{118}{3}\frac{M_{\tilde{g}}}{m_q}
~ \widetilde{C}^q_5(M_S)~.
 \label{initial_FV2}
\end{align}
These initial conditions as well as the RGE \eqref{rge} are again
consistent with the one-loop results given in
Ref.~\cite{Hisano:2008hn}. 

\begin{figure}[t]
\begin{center}
\includegraphics[width=70mm]{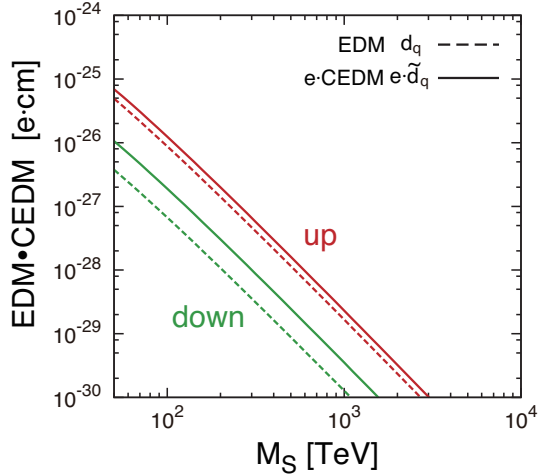}
\caption{Quark EDMs $\vert d_q\vert$ and CEDMs $e\vert\tilde{d}_q\vert$
 at the hadron scale 
 $\mu_H=1~{\rm GeV}$ as functions of $M_S$. Solid and dashed lines
 represent the CEDMs and EDMs, respectively. Upper two red lines are for
 up quark, while lower two green lines for down quark. We take
 $M_{\tilde{g}}=3~{\rm TeV}$, $\tan\beta=3$, $\vert \mu\vert =M_S$, and
 $A_q=0$. Mass insertion parameters and phase factor are assumed to be $\vert
(\delta_{LL})_{qq_3^{}}\vert=\vert (\delta_{RR})_{q_3^{}q}\vert=1/3$ and
$\sin\theta_q=1/\sqrt{2}$, respectively.} 
\label{edmcedm}
\end{center}
\end{figure}

By using a similar procedure to that described in the previous section,
we readily evaluate the EDMs and CEDMs at the gluino mass scale with
initial conditions \eqref{initial_FV} and \eqref{initial_FV2}. 
Let us now evolve them down to the hadron scale.
Below the gluino threshold, the gluino fields are integrated out and the
effective theory includes only the SM fields. The tree-level matching
condition is applied to $C^q_1$ and $C^q_2$, and then they are evolved
down to the hadronic scale in terms of the SM RGEs. The quark EDMs and
CEDMs are then given as
\begin{align}
 d_q&=m_q(\mu^{}_H)eQ_qC^q_1(\mu^{}_H)~,\nonumber\\
 \tilde{d}_q&=m_q(\mu^{}_H)C^q_2(\mu^{}_H)~,
\end{align}
with $\mu^{}_H\sim 1$~GeV the hadron scale.

In Fig.~\ref{edmcedm}, the absolute values of the quark EDMs $\vert
d_q\vert$ and CEDMs $e\vert\tilde{d}_q\vert$ at the 
hadron scale  $\mu^{}_H=1~{\rm GeV}$ are plotted as functions of the squark
mass scale $M_S$. The solid and dashed lines represent the CEDMs and
EDMs, respectively. The upper two red lines correspond to the EDM and
CEDM of up quark, while the lower two green lines to those of down
quark. Here, we take $M_{\tilde{g}}=3~{\rm TeV}$,
$\tan\beta=3$, $\vert \mu\vert =M_S$, and $A_q=0$.\footnote{In the
anomaly mediation, the $A$-terms are suppressed by one-loop factors,
thus negligible in our calculation.} In addition, the mass insertion
parameters and the phase factor are assumed to be $\vert
(\delta_{LL})_{qq_3^{}}\vert=\vert (\delta_{RR})_{q_3^{}q}\vert=1/3$ and
$\sin\theta_q=1/\sqrt{2}$ with $\theta_q\equiv {\rm
Arg}[\mu(\delta_{LL})_{qq_3^{}}(\delta_{RR})_{q_3^{}q}]$, respectively. From
this figure, we find that the CEDMs dominate the EDMs, though the latter are
not negligible at all. Further, the contribution of up quark is larger
than that of down quark in the case of low $\tan\beta$, which is favored
from the viewpoint of the 126~GeV Higgs mass in the high-scale SUSY
scenario \cite{Giudice:2011cg, Ibe:2011aa, Ibe:2012hu, Ibanez:2013gf}.
We would like to remark that the
EDMs and CEDMs are proportional to the gluino mass except for the
renormalization factors, and thus their values corresponding to other
gluino masses are readily obtained by means of the scaling law, as long
as $M_{\tilde{g}}\ll M_S$.

\begin{figure}[t]
\begin{center}
\includegraphics[width=70mm]{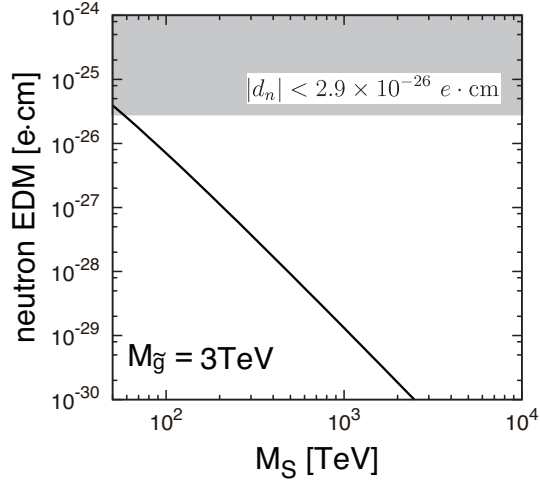}
\caption{Neutron EDM $d_n$ as a function of $M_S$. The same parameters
 are used as those exploited in Fig.~\ref{edmcedm}. Shaded region
 represents the current experimental limit $\vert d_n\vert <2.9\times
 10^{-26}~e\cdot{\rm cm}$ \cite{Baker:2006ts}. } 
\label{nEDM}
\end{center}
\end{figure}

By using the EDMs and CEDMs computed above, we finally calculate the
neutron EDM $d_n$. To that end, we need to express the neutron EDM 
in terms of $d_q$ and $\tilde{d}_q$. At present, only the
calculations based on the QCD sum-rules \cite{hep-ph/0010037,
Hisano:2012sc} include both of these contributions on an equal
footing. Their theoretical error is, however, still significant, though
partial use of lattice results for the low-energy QCD constants may
reduce the uncertainty \cite{Hisano:2012sc}. Moreover, this approach
seems to lack the strange quark contributions. For instance, when
one imposes the Peccei-Quinn symmetry, the strange CEDM contribution to
the neutron EDM completely vanishes in the case of the sum-rule
calculations, while it is expected to be sizable from the estimation
based on the chiral perturbation theory \cite{Fuyuto:2012yf}.
At this moment, both methods have large uncertainty and no consensus has
been reached yet. We strongly anticipate that the lattice
simulations will evaluate the neutron EDM induced by the quark EDMs and
CEDMs with high accuracy. In the present calculation, we use the result
presented in Ref.~\cite{Hisano:2012sc}:\footnote{The
numerical values presented here are in fact different from those
in Ref.~\cite{Hisano:2012sc} by nearly a factor of two. The difference
results from the use of different values for the quark condensate; We
use  $\langle
\bar{q}q\rangle=-m_{\pi}^2f_{\pi}^2/(m_u+m_d)\simeq-(262~{\rm MeV})^3$ \cite{PDG2013} while $\langle \bar{q}q\rangle=-(225~{\rm MeV})^3$ is used in
Ref.~\cite{Hisano:2012sc}. }
\begin{equation}
 d_n=0.79 d_d-0.20d_u+e(0.30\tilde{d}_u+0.59\tilde{d}_d)~,
\label{dnPQ_sum_rule}
\end{equation}
where we assume the Peccei-Quinn mechanism. In
Fig.~\ref{nEDM}, we plot the resultant neutron EDM as a function of
$M_S$. In this figure, we use the same parameters as those exploited
in Fig.~\ref{edmcedm}. The shaded region represents the current
experimental limit $\vert d_n\vert <2.9\times 10^{-26}~e\cdot{\rm cm}$
\cite{Baker:2006ts}. As seen from this figure, the present experimental
limit has already excluded the squark mass scale nearly up to
$10^2$~TeV. Future experiments of the neutron EDM are expected to reach
$\sim 10^3$~TeV, which covers most of the region favored from the
high-scale SUSY scenario compatible with the 126~GeV Higgs mass and the
existence of ${\cal O}(1)$~TeV gauginos. Hence, the EDM experiments are
quite promising, and may be about to grasp the signature of
supersymmetry. 

In the case of the minimal flavor violation discussed in the previous
section, on the other hand, the predicted neutron EDM lies around
$d_n\simeq10^{-30}~e\cdot{\rm cm}$ for $M_S=10^2$~TeV, which is much
below the current experimental limit.

\section{Conclusion and discussion}
\label{conclusion}

In this paper, we discuss the QCD corrections to the dimension-five
CP-violating operators in
the case of the high-scale supersymmetry. To appropriately evaluate the
radiative corrections in the presence of a large hierarchy between the
squark and gluino mass scales, we exploit the RGEs 
(including CP violating gluino-quark four-Fermi operators)
in an effective
theory where only the SM particles and gluinos are taken into
account. As a result, the values of the low-energy quark EDMs and CEDMs
may differ from those evaluated in previous works by ${\cal O}(100)$~\%
and ${\cal O}(10)$~\%, respectively.

In the high-scale SUSY scenario, similar calculations based on
the RGEs may have significant consequences for the prediction of 
other low-energy observables, such as gluino decay rates
\cite{Gambino:2005eh, Sato:2012xf, Sato:2013bta},
particle-antiparticle mixing, rare and CP-violating decays, and so on.  
Even though these processes are often induced by the flavor-changing
operators, a lot of our results are applicable to the cases since
gluinos as well as photons and gluons do not distinguish quark flavors.

In the above calculation, we have only included the leading order
effects, though the one-loop short-distance correction is also
discussed. Before concluding this article, let us discuss possible ways
of improvement of the above calculation. A straightforward
improvement is achieved if one uses the two-loop RGEs as well as
a complete set of one-loop threshold corrections. In addition, to go beyond
the leading order analysis, we also need to include the operators which
we neglect in our calculation; the gluino CEDM, four-quark
operators, four-gluino operators, and the Weinberg operator. These
operators mix with each other as well as with the quark EDMs and CEDMs
during the RGE flow. A complete calculation beyond the leading order
will be carried out on another occasion \cite{FHNT}.

~\\~\\
{\it Note Added:} While this work was being finalized, we realized the
authors in Ref.~\cite{Altmannshofer:2013lfa} estimated the
anomalous dimensions for the quark-gluino four-Fermi operators in a
similar context. The results presented in the reference are, however,
inconsistent with ours. Especially, the authors
insist that they have not found the mixing among the
four-Fermi operators by their explicit calculation, though we do as
shown in Eq.~\eqref{gammag}.

\section*{Acknowledgments}

The work of J.H. is supported by Grant-in-Aid for Scientific research from the
Ministry of Education, Science, Sports, and Culture (MEXT), Japan,
No. 24340047, No. 23104011 and No. 22244021, and also by
World Premier International Research Center Initiative (WPI
Initiative), MEXT, Japan.
The work of N.N. is supported by Research Fellowships of the Japan Society
for the Promotion of Science for Young Scientists.

{}

\end{document}